# Haptics-Augmented Physics Simulation: Coriolis Effect

**Felix G. Hamza-Lup, Benjamin Page**

Computer Science and Information Technology
Armstrong Atlantic State University
Savannah, GA 31419, USA
E-mail: felix.hamza-lup@armstrong.edu

**Abstract**

*The teaching of abstract physics concepts can be enhanced by incorporating visual and haptic sensory modalities in the classroom, using the correct perspectives. We have developed virtual reality simulations to assist students in learning the Coriolis effect, an apparent deflection on an object in motion when observed from within a rotating frame of reference. Twenty four undergraduate physics students participated in this study. Students were able to feel the forces through feedback on a Novint Falcon device. The assessment results show an improvement in the learning experience and better content retention as compared with traditional instruction methods. We prove that large scale deployment of visuo-haptic re-configurable applications is now possible and feasible in a science laboratory setup.*

**Introduction**

Haptic interfaces allow us to touch and interact with virtual objects simulated on a computer as if they were real. In addition to feeling virtual objects' properties, virtual forces such as gravity, friction, and tension may be simulated as well. The advent of haptic technology has made affordable haptic hardware interfaces widely available. Furthermore, open-source APIs allow for rapid prototyping of visuo-haptic simulations.

The Coriolis effect, named for the French mechanical engineer Gaspard-Gustave de Coriolis (1792-1843), is a perceived force that alters the path of a moving object, depending on the hemisphere. For example, an airborne object travelling away from the North Pole would appear to be forced to the right, or West, due to the rotation of the Earth. Contrarily, an airborne object travelling from the South Pole would appear to be forced to the left, or East. This force is described as being *perceived* because the observer's rotating frame of reference creates the phantom force that acts upon the object (Stansfield, 2009); however, to an observer from a fixed from of reference, no force acts upon the object. When travelling on the Earth's surface, an object will appear to move in a straight line, when in reality its path is curved as the Earth rotates.

Observing the Coriolis effect in the real-world is frequently done using rotating carousels or children's merry-go-rounds. When viewed from a fixed vantage point above, a ball thrown from the center of a rotating object will appear to move in a straight line. When viewed from a rotating vantage point, such as while standing on the edge of a carousel, the ball would appear to curve in an arc (Ehrlich, 1990).

A need for a practical, hands-on approach to teaching this concept arose, as the concept is difficult to describe concisely with text and still images. Fieldtrips outside the classroom are often impractical due to time and cost. In cases where a fieldtrip is possible, witnessing the effect while standing on a rotating surface can induce nausea from motion sickness (Ginns, 2006).

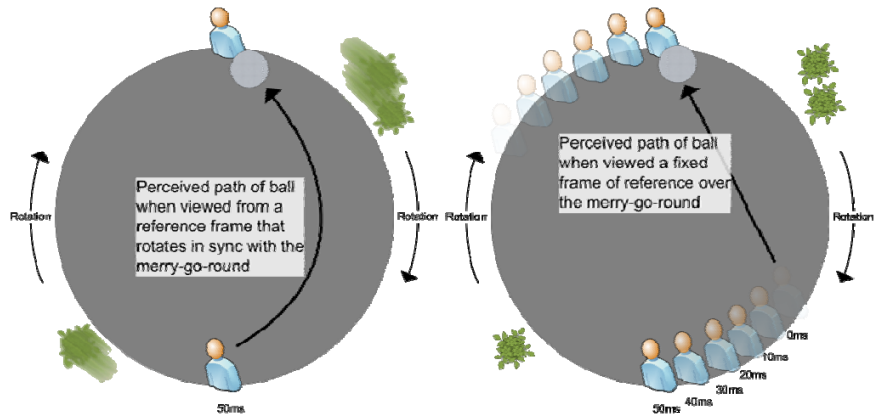
Figure 1.**The Coriolis Effect (Left) Vs. Fixed Reference Frame (Right)**

Our goal in developing a visuo-haptic simulation for the Coriolis effect was to enable physics students to observe and feel the motion of an object and the forces acting upon it while moving over a rotating frame of reference, whether airborne or at the surface. We evaluate the effectiveness of using active haptics with force-feedback, haptic hardware without force-feedback, and traditional classroom approaches. By doing so, we hope to find novel instructional methods that will best allow students to achieve a better understanding of the forces, specifically the abstract forces responsible for the Coriolis effect, at work in a rotating environment.

**Background: Hardware and Software**
Haptics3D is a well-known open source API. H3D API provides a link between the graphics and haptics rendering, allowing haptic devices to interact with 3D rendered objects. The main advantages of H3D are the rapid prototyping capability and the compatibility with X3D, making it easy for the developer to manage both the 3D graphics and the haptic technology from SensAble's OpenHaptics$^{TM}$ toolkit. It allows users to focus their work on the behaviour of the application, and ignore the issues related to haptics geometry rendering. The API is also extended with scripting capabilities, allowing the user to perform rapid prototyping using the Python scripting language. The Python scripts in an H3D environment contain the logic to control the properties of nodes referenced from the X3D files. By binding routes to and from X3D objects, events may be handled programmatically.

The haptic hardware used in this study is a Novint Falcon controller, a consumer grade haptics 3D mouse with a 4"x4"x4" working volume and sub-millimeter resolution, 1ms per instruction, and a maximum force of approximately two pounds in any direction (Ogando, 2007).

**Simulator Interface**
Our goal in developing the simulation was to target multiple sensory modes (visual and haptic), as multimodal learning increasing sensory bandwidth, and has been shown allow for faster learning, with greater retention, than with traditional teaching methods alone (Jones, 2006)(Bara et al, 2007).

We created a simulation where students could push a ball with friction from a rotating surface, and observe this from a vantage point that rotated at the same rate as the surface. The path of the ball is also drawn so the student is able to see if the ball moves straight or curves, as shown in Figure 2. This allowed students to feel the phantom force of the Coriolis effect.

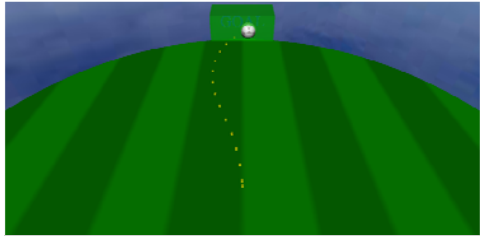
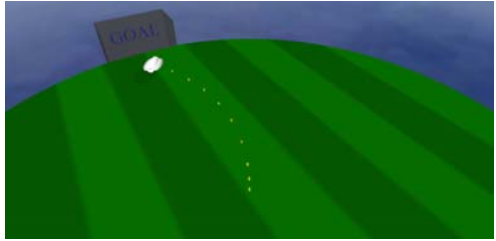

Figure 2. **Coriolis Effect Ball Simulation**                    Figure 3. **Glider Simulation**

In order to allow students a juxtaposition between a rotating vantage point versus a fixed vantage point, we developed a second demo with a glider moving without surface friction, from the perspective of a fixed vantage point, as shown in Figure 3. From a fixed perspective, the lack of a phantom force could be observed.

**Experimental Design**
The participants of this experiment were 24 undergraduate college students taking Principles of Physics I at Armstrong Atlantic State University in Savannah, Georgia. The students were divided into four groups of six students, based on their GPA such that each group's average GPA and GPA variance (average square deviation from the mean) was similar. The four groups were divided as follows:

- Group 1 –Presented with supplemental reading material and a video on the Coriolis effect before filling out a questionnaire.
- Group 2 – This group was given the same supplemental reading material and video, then participated in a visual simulation with no haptic feedback, followed by a questionnaire.
- Group 3 – Also was given the reading material and video, then participated in a visuo-haptic simulation involving force feedback, followed by a questionnaire.
- Group 4 – Also was given the reading material and video, then was given a tutorial on using the haptic devices to become familiar with the hardware, then participated in a visuo-haptic simulation with force feedback, followed by a questionnaire.

Table 1 below shows the group pairs, the independent variables observed, and the dependent variables for each pair as we try to find if a correlation exists between the proposed variables. Objective assessment is done by comparing students' quiz results, and subjective assessment was done through student answers in a questionnaire.

Table 1. **Group Pairs for Each Experiment**

| Pair ID | Control group | Experimental group | Independent variable | Dependent variable |
|---------|---------------|--------------------|-----------------------|--------------------|
| G1-G4   | Group 1       | Group 4            | Trials and visuo-haptic simulation | Quiz Score |
| G2-G3   | Group 2       | Group 3            | Haptic component      | Quiz Score |
| G3-G4   | Group 3       | Group 4            | Trials                | Quiz Score |
| G1-G3   | Group 1       | Group 3            | Visuo-haptic simulation | Quiz Score |

**Results**

To evaluate the performance of the simulations, a combined quiz score for each group of students was calculated by combining their individual scores. The groups' quiz results are shown in the Figure 4.

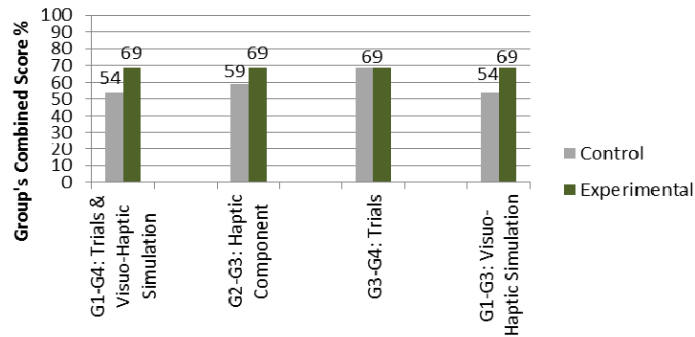

Figure 4. **Quiz Score Comparisons**

Both the G1-G4 and the G1-G3 pairing show a 15% increase in quiz scores for the groups that took part in a visuo-haptic simulation of the Coriolis effect, versus the group that only had reading material and a video. The G2-G3 paring shows a 10% increase in quiz scores for the group that participated in a simulation without haptic feedback. There was no difference in quiz scores in the G3-G4 pairing, showing that a tutorial on the haptic hardware prior to the Coriolis effect simulations either didn't sufficiently increase students' familiarity, or they familiarized quickly to the simulations and practice was not necessary.

   Feedback was collected from the students in order to gain an understanding of their perceptions of the effectiveness of the physics simulations. Most students were observed having difficulty controlling the ball and aircraft with the haptic device; however, once students understood the controls, they reported that it felt 'natural' and 'quite simple'. 94% of the students had positive comments on the effectiveness of the simulation. A frequent comment among students suggested that having hands-on, interactive simulations allowed them to understand the forces at work better than simply reading about them or watching lectures. A student from Group 4 (force feedback simulation and 10 practice trial) had this to say, "It was an interesting experience and a good way to gain perspective. It wasn't difficult at all to use the device, especially with the tutorial exercise beforehand. It would absolutely be beneficial to use the device again."

**Conclusion**

Visuo-haptic simulation is particularly interesting for scientific applications where haptics, combined with 3D visualization, may provide accurate and rapid understanding of concepts such as abstract physics phenomena.

   As we observed from experiments over the last few years, there are additional advantages to such simulations: repeatability of the experiments and a large (effectively continuous) range of physical parameters that can be customized by the user for a particular experiment.

   Cognitive studies have shown that students would engage with learning material if they could easily understand abstract/difficult concepts and relate new information to what they already know. Lack of attention and engagement, however, results in more failing grades, more expulsions, increased dropout rates, and a lower rate of undergraduate completion, especially in STEM disciplines. We are proposing a new and deeper level of engagement, which (from our preliminary assessment) significantly and consistently improves student interest in the subject matter.

This research shows that by augmenting traditional classroom learning with the use of haptic tools, students were able to better understand abstract physics concepts. This was reflected in higher quiz scores for the groups who participated in the simulations. For future research, we would like to explore long term retention of knowledge gained from visuo-haptic simulations. We would also like to explore visuo-haptic methods of instruction in additional disciplines.